\documentstyle[12pt]{article}

\begin{document}
\begin{center}
\large
{\bf Conditional Hamiltonian and Reset Operator\\
in the Quantum Jump Approach}
\footnote{To appear in {\em Quantum and
Semiclassical Optics}}
\\[.8cm]
Gerhard C. Hegerfeldt\cite{email}  and Dirk G. Sondermann
\end{center}
\normalsize
\quad
\begin{center}
Institut f\"ur Theoretische Physik, Universit\"at G\"ottingen\\
Bunsenstr. 9, D-37073 G\"ottingen, Germany \\[1.cm]
\end{center}
\begin{abstract}
{For the time development of a single system in the quantum jump
approach or for quantum trajectories one requires the conditional
(reduced) Hamiltonian between jumps and the reset operator
after a jump. Explicit expressions for them are derived for a general
$N$-level system by employing the same assumptions as in the usual
derivation of the Bloch equations.
We discuss a possible minor problem with positivity for these
expressions as well as for the corresponding Bloch equations.\break
}
\end{abstract}
\vspace*{1cm}

\noindent {\bf 1. Introduction}\\[0.5cm]
In order to better understand photon statistics and photon counting
processes for single atoms, in particular the macroscopic dark
periods (``electron shelving") in the Dehmelt $V$ system \cite{dark},
the quantum jump approach was developed in G\"ottingen prior to 1991
\cite{HeWi,reset}, and consequently analytic aspects were stressed. The
independently proposed Monte-Carlo wave function approach \cite{Dal}
is essentially equivalent, but stressed more the computational
aspects. The approach by quantum trajectories \cite{Car} was also
independently proposed and is also essentially equivalent. There are
various papers which can be regarded as precursors to the quantum
jump approach, in particular Refs. \cite{Mollow,MarteZoller}. Some
related  work \cite{PP,Reibold,DS} on photon  statistics
will be discussed in Section 5.

Our analytic methods have been applied to interference effects in
single atoms \cite{HePl1,HePl2,HePl3,HePl4}. Recently, the
quantum jump approach has been extended to study conditional
fluorescence spectra, such as the spectrum of an atom in an extended
light period \cite{HePl6}.

In the derivation of the quantum jump approach one envisages photon
measurements in rapid succession at times $\Delta t$ apart. If a
photon is detected at some time it is, for simplicity, assumed to be
absorbed. This assumption, however, is not essential \cite{Pl}.
$\Delta t$ should be much smaller than the level life-times, but it
cannot be too small because then one would encounter the quantum Zeno
effect \cite{QZE}; therefore $\Delta t$ should be larger than the
inverse optical frequencies. A reasonable range is $\Delta t \approx
10^{-12}$s.

Between broadband photon detections the time development of an atom is
then described by a reduced, or rather conditional, Hamiltonian
$H_{{\rm cond}}$, the condition being that no photons are found. It is
non-hermitian and allows the calculation of the detection or emission
probability. Once a photon is detected, the atomic state has to be
reset  (a ``jump"), usually to the ground state, but for general
$N$-level systems the reset states may be quite complicated \cite{reset}.
Thus the state of a single atom is given by a random path (``quantum
trajectory") determined by the conditional Hamiltonian $H_{{\rm cond}}$
and the reset states.

In Ref. \cite{reset}  one of us derived a simple expression for the reset
state in the two limiting cases of either widely separated or closely
neighbored transition frequencies. The general case seemed to require
much smaller $\Delta t$'s, which are difficult to justify physically.

The main purpose of this paper is to re-investigate the reset state
using the same methods and approximations as in the derivations of
the Bloch equations \cite{Mi,PP}, mainly the Markov approximation.
It will be shown that in the general case the (super-)operator which
resets the state may depend in a complicated oscillatory way on the
choice of $\Delta t$. However, this ``physical" reset operator can be
replaced by a simpler refined or idealized  reset operator which
gives the same results for all photon counting processes. This
idealized reset operator turns out to have the same form as the one
obtained in Ref. \cite{reset} by using very small $\Delta
t$'s. This is explained in detail in Sections 3, while in
Section 2 a short derivation of the conditional (or reduced) Hamilton
is given. In Section 4 it is shown that the collection of quantum
trajectories satisfies the Bloch equations, both for the physical and
for the idealized reset operator. In the last section we discuss our
results and the positivity of the expressions. It turns out that in
some cases
the reset operator and the general damping matrix in the conditional
Hamiltonian can have a small negative part.
A related problem  also occurs in  Bloch equations
where in some examples the solution develops a small negative part for
very small times  and  thus temporarily looses its positivity
\cite{Haake}. But this is probably of no practical relevance.
\\[1cm]
\noindent {\bf 2. The conditional Hamiltonian $\bf H_{\bf cond}$ for
a general N-level system}\\[0.5cm]
With an external field ${\bf E}_e(t)$, the Hamiltonian for
complete system, atom plus quantized electromagnetic field ${\bf E} =
{\bf E}^{(+)} + {\bf E}^{(-)}$, is given in the Schr\"odinger
picture, in the limit of long wave-lengths and in the rotating-wave
approximation, by \cite{Lou}
\begin{eqnarray}
H & = &
\sum \hbar \omega_i \left|i\right>\left<i\right| + H^0_F +
\Bigl[e {\bf D}^{(-)} \cdot {\bf E}_e^{(+)} (t) + \hbox{h.c.}\Bigr] +
\Bigl[e {\bf D}^{(-)} \cdot {\bf E}^{(+)} + \hbox{h.c.}\Bigr]\nonumber\\
& \equiv & H^0_A + H^0_F + H_{AL}(t) + H_{AF} \label{1}
\end{eqnarray}
where
\begin{eqnarray}
{\bf D}^{(-)} & = & \sum_{i > j}{\bf D}_{i j}
|i\rangle \langle j| \nonumber\\
{\bf D}_{i j} & = & \langle i | {\bf X} |j \rangle \nonumber\\
{\bf E}^{(+)} & = & \sum_{{\bf k} \lambda} i e \left\{ \frac{\hbar
\omega_{\bf k}}{2 \varepsilon_0 V} \right\}^{1/2} {\bf \varepsilon}_{{\bf k}
\lambda} a_{{\bf k} \lambda}. \label{3}
\end{eqnarray}
and a frequency cutoff is included.
$V$ is the quantization volume, later taken in infinity, and
$i>j$ means $\omega_i>\omega_j$.

Now one imagines that photon measurements are performed at times
$t_1 = \Delta t , t_2 = 2 \Delta t, \cdots$ on an ensemble of systems
which, at time $t = 0$, is supposed to be in the initial state $|
0_{ph} \rangle | \psi \rangle$. By the von Neumann - L\"uders
projection postulate \cite{N} the {\em subensemble} for which no
photons are detected until time $t_n $ is described, with $ P_0
\equiv | 0_{ph} \rangle {\bf 1}_A \langle 0_{ph} |$, by
\begin{equation}\label{4}
   P_0 U(t_n, t_n - \Delta t) P_0 \cdots P_0 U(t_1, 0) | 0_{ph}
\rangle | \psi \rangle \equiv | 0_{ph} \rangle U_{{\rm cond}} (t_n, 0)
| \psi \rangle~.
\end{equation}
The norm squared of this is the probability of finding no photons for
the measurements between $0$ and $t_n$. $U_{{\rm cond}}$ gives the time
development of an atom under the condition that no photon is observed
until time $t$, and it will now be determined explicitly by simple
second order perturbation theory. Going over to the interaction
picture with respect to $H_A^0 + H^0_F$ one has
\begin{equation}\label{5}
H_I (t) = H_{AL}^I (t) + H^I_{AF} (t)
\end{equation}
which is obtained by replacing $| i \rangle \langle j |$
and $a_{{\bf k} \lambda}$ in the original interaction Hamiltonian
by $| i \rangle \langle j | e^{i \omega_{i j} t}$ and
$a_{{\bf k} \lambda} e^{-i \omega_{\bf k}t}$, respectively, with
$$ \omega_{ij}\equiv\omega_i-\omega_j \,. $$
We now calculate, for $t_i\leq t'< t_{i+1}$,
\begin{equation}\label{7}
\langle 0_{ph} | {d\over dt'} U_I (t',t_i) | 0_{ph} \rangle~.
\end{equation}
In the first-order contribution only $H_{AL}^I (t)$ remains since
\begin{equation}\label{8}
\langle 0_{ph} | H_{AF}^I (t) | 0_{ph} \rangle = 0~.
\end{equation}
The second order is, by Eq. (\ref{8}),
\begin{equation}\label{9}
- \hbar^{-2} \int^{t'}_{t_i} d t''  \left\{
\langle 0_{ph} | H_{AF}^I (t') H^I_{AF} (t'') | 0_{ph} \rangle
+ H^I_{A L} (t') H^I_{AL} (t'' ) \right\}~.
\end{equation}

Now, if the external field ${\bf E}_e (t)$  is smooth in time, e.g.
given by laser, and {\em not} wildly fluctuating like a thermal or
chaotic field, then the second part in Eq. (\ref{9}) contributes a
term of higher order in $\Delta t$ and can therefore be omitted. For a
chaotic external field, however, this part may give rise to a
contribution of the same order and then has to be retained.
In this
case one can no longer work with state vectors (``wave functions")
but has to use (conditional) density matrices. A particular example
of this is treated in Ref. \cite{HePl2}.

Thus, supposing a smooth external field, the second-order
contribution becomes
$$ - \hbar^{-2}  \int_{t_i}^{t'} dt''
   \sum_{ij \ell m \atop i>j,~\ell>m}
   |i\rangle\langle j|m\rangle\langle\ell|
   \sum_{{\bf k} \lambda}
   \frac{e^2 \hbar \omega_k}{2 \varepsilon_0 V}
   ({\bf D}_{i j} \cdot {\bf \varepsilon}_{{\bf k} \lambda})
   ({\bf \varepsilon}_{{\bf k} \lambda} \cdot {\bf D}_{m \ell})
   e^{-i(\omega_{\bf k} - \omega_{  i  j}) t'+
       i(\omega_{\bf k} - \omega_{\ell m}) t''} $$
\begin{equation}\label{10}
 =  - \hbar^{-2}  \sum_{ij\ell \atop i,\ell>j}
   e^{i(\omega_{ij} - \omega_{\ell j})t'}
   |i\rangle\langle\ell|
   \int^{t'-t_i}_0 d\tau \sum_{{\bf k} \lambda}
   \frac{e^2 \hbar \omega_k}{2 \varepsilon_0 V} ({\bf D}_{i j} \cdot {\bf
   \varepsilon}_{{\bf k} \lambda})({\bf
   \varepsilon}_{{\bf k} \lambda} \cdot {\bf D}_{j \ell})
   e^{-i(\omega_{\bf k} - \omega_{\ell j} )\tau}~.\end{equation}

One can now use properties of the correlation function
\begin{equation} \label{10a}
\kappa_{ji\ell m}(\tau)  \equiv \sum_{{\bf k} \lambda}
\frac{e^2 \omega_{\bf k}}{2 \varepsilon_0 \hbar V} ({\bf D}_{ji} \cdot
{\bf \varepsilon}_{{\bf k} \lambda})
({\bf \varepsilon}_{{\bf k} \lambda} \cdot
{\bf D}_{\ell m})~e^{-i(\omega_{\bf k} - \omega_{\ell m})\tau}~.
\end{equation}
With $V^{-1} = \Delta^3 k / (2 \pi)^3$
one can  perform the limit $V \rightarrow \infty$, and the sum over
${\bf k}$ becomes an integral over $\omega$, with
a suitable frequency cutoff, and an integral over the unit sphere.
The correlation function has an effective width of the order of
$\omega^{-1}_{\ell m}$ around $\tau = 0$, and for $t'-
t_i \gg \omega^{-1}_{\ell m}$ one can therefore extend the $\tau$
integration in Eq. (\ref{10}) to infinity \cite{width}.
This is equivalent to the approximation
\begin{equation}\label{11}
\int^{t'-t_i}_0 d \tau e^{i(\omega_{\ell m} - \omega_{{\bf k}}) \tau} \cong
\pi \delta (\omega_{\bf k} - \omega_{\ell m}) + i {\cal P}
\frac{1}{\omega_{\bf k} - \omega_{\ell m}}
\end{equation}
and corresponds to the usual Markov approximation in the derivation
of the Bloch equations \cite{Mi,PP}.
For the second-order contribution one then obtains
\begin{equation}\label{11a}
- \sum_{{i j \ell \atop i, \ell > j}} | i \rangle \langle \ell |
e^{i \omega_{i \ell} t'}  \int d^3 k
\frac{e^2 \omega_{\bf k}}{(2 \pi)^3 \hbar 2 \epsilon_0} {\bf D}_{i j}
\cdot \sum^2_{\lambda = 1} {\bf \varepsilon}_{{\bf k} \lambda} {\bf
\varepsilon}_{{\bf k} \lambda} \cdot {\bf D}_{j \ell}
(\pi \delta(\omega_{\bf k} - \omega_{\ell m})~
+ i {\cal P}\frac{1}{\omega_{\bf k} - \omega_{\ell m}})~.
\end{equation}
The principal-value term is analogous to a level shift and is often
omitted \cite{PP,CT}. The last integral equals $\Gamma_{i j j \ell}$ where
\begin{equation}\label{12}
\Gamma_{i j k \ell} \equiv \frac{e^2}{6 \pi \varepsilon_0 \hbar c^3}
{\bf D}_{ij}\cdot {\bf D}_{k \ell} | \omega_{k \ell} |^3~~~
+~~~\mbox{principal value term}
\end{equation}
Hence, integrating over $t'$  from $t_i$ to $t_{i + 1}$ and
using $1 + \delta \approx e^\delta$ for small $\delta$, we obtain
\begin{equation} \label{12a}
\langle 0_{ph} | U_I (t_{i + 1}, t_i) | 0_{ph} \rangle = \exp \left\{
- \frac{i}{\hbar} \int^{t_{i + 1}}_{t_i} dt' \left\{ H_{AL} (t') - i
\hbar \sum_{ij \ell \atop i,\ell > j} \Gamma_{i j j \ell} e^{i
\omega_{i \ell}t'}
|i \rangle \langle \ell| \right\} \right\}~.
\end{equation}
For small $\Delta t$ this can be replaced by a time-ordered
exponential and thus, with $t = t_n,$
\begin{equation}\label{13}
\prod^{n~~}_{~~~1} \langle0_{ph}| (U_I (t_{i}, t_{i-1}) |
0_{ph} \rangle \cong {\cal T} \exp \left\{ - \frac{i}{\hbar} \int^t_0
dt' \left\{ H_{AL} (t') - i \hbar \sum_{ij \ell \atop
i,\ell > j} \Gamma_{i j j
\ell} e^{i \omega_{i \ell} t'} | i \rangle \langle \ell | \right\}
\right\}
\end{equation}
where the product sign on the l.h.s. includes an ordering in an
obvious way.

Since
\begin{equation}\label{14}
\langle 0_{ph} | U(t_i, t_{i-1}) | 0_{ph} \rangle = e^{-i H^0_A t_i
 / \hbar} \langle 0_{ph} | U_I (t_i, t_{i-1}) | 0_{ph} \rangle
e^{i H^0_A t_{i-1} / \hbar}
\end{equation}
and since, for $t = t_n = n \Delta t$,
\[
U_{{\rm cond}}(t,0) = \prod^{n~~}_{~~~1} \langle 0_{ph}
|U (t_i, t_{i-1}) | 0_{ph} \rangle,
\]
we obtain, on a coarse-grained time scale, from Eqs. (\ref{13}) and
(\ref{14})
\begin{equation}\label{15}
U_{{\rm cond}} (t,0) = {\cal T} \exp \left\{ - \frac{i}{\hbar}
\int^t_0 dt' \left\{ H^0_A + H_{AL}(t') - i \hbar
\sum_{{i \ell j \atop i, \ell > j}} \Gamma_{i j j \ell} | i \rangle
\langle \ell | \right\} \right\}
\end{equation}
which is the transformation of Eq. (\ref{13}) back to the
Schr\"odinger picture.

Thus, with the atomic  operator $\Gamma$ defined as
\begin{equation}\label{16}
\Gamma \equiv \sum_{{i \ell j \atop i, \ell > j}} \Gamma_{i j j \ell}
| i \rangle \langle \ell |
\end{equation}
the conditional Hamiltonian for an $N$-level atom with no photon
emission until time $t$ is, on the coarse-grained time scale,
given by
\begin{equation}\label{17}
H_{{\rm cond}}(t) = H^0_A + H_{AL}(t) - i \hbar \Gamma~.
\end{equation}

For initial atomic state $| \psi \rangle$ the probability to find  no
photon until time $t$ is thus given by

\begin{equation}\label{17a}
   \| U_{{\rm cond}} (t, 0) | \psi \rangle \|^2~,
\end{equation}
and the probability to find the first photon in $(t, t + \Delta t)$
is the difference of this expression for $t$ and $t + \Delta t$.
Thus, on the coarse grained time scale the probability density $w(t)$
for the first photon is the negative derivative of Eq. (\ref{17}),
\begin{eqnarray}
w (\tau) & = & -\frac{d}{d t} || U_{{\rm cond}} (t, 0) | \psi
\rangle ||^2\nonumber \\
& = & \langle \psi | \Gamma + \Gamma^* | \psi \rangle~.
\label{17b}
\end{eqnarray}

When one lets $\Delta t$ become smaller and smaller
the above derivation shows very nicely how and where the quantum Zeno
effect \cite{QZE} turns up in a very natural way.If $\Delta t$ is
chosen much smaller than the inverse optical frequencies, the last
exponential in Eq. (\ref{10}) can be replaced by 1, and the integral
becomes proportional to $t'-t_i$.Eq. (\ref{12a}) is then replaced by
\begin{equation} \label{12aa}
\langle 0_{ph} | U_I (t_{i + 1}, t_i) | 0_{ph} \rangle = \exp \left\{
- \frac{i}{\hbar} \left\{ H_{AL} (t_i)\Delta t - i \hbar~ {\rm const}~
\sum_{ij \ell \atop i,\ell > j} \Gamma_{i j j \ell} e^{i \omega_{i
\ell}t_i}|i \rangle \langle \ell| (\Delta t)^2
\right\} \right\}~.
\end{equation}
The product of these operators then becomes, for $\Delta t
\rightarrow 0$,
\begin{equation}\label{13aa}
 {\cal T} \exp \left\{ - \frac{i}{\hbar} \int^t_0
dt' \left\{ H_{AL} (t')  \right\}\right\}~.
\end{equation}
This is a purely atomic operator, and hence the time development
of the field becomes frozen, i. e. for $\Delta t \rightarrow 0$ one
always remains in the vacuum. For this reason one cannot choose
$\Delta t$ arbitrarily small in the quantum jump approach.
\vspace*{1cm}

\noindent {\bf 3. The reset operator}

\vspace*{0.5cm}

\noindent In this section we determine the state (or density matrix)
of an atom after a broadband detection of a photon, under the
condition that shortly before no photon was found. We therefore
consider an ensemble where no photon are present at time $t_n$. The
ensemble is thus described by $\rho (t_n) = | 0_{ph} \rangle
\rho_A(t_n) \langle 0_{ph} |$ where $\rho_A$ is the density matrix of
the atoms. The state of the subensemble for which photons are
detected by a non-absorptive measurement
at time $t_{n + 1} = t_n + \Delta t$ is
given, in view of the von Neumann-L\"uders projection postulate
\cite{N}, by
\begin{equation}\label{18}
P_1 \rho (t_{n + 1}) P_1 / \hbox{tr} (\cdot)
\end{equation}
where
\begin{equation}\label{19}
P_1 \equiv {\bf 1} - | 0_{ph} \rangle {\bf 1}_A \langle
0_{ph} |~.
\end{equation}
Note that Eq. (\ref{18}) still  contains the photons.

After a photon measurement by absorption no photons are present any
longer and it was argued in Ref. \cite{reset} that the resulting
reset state is obtained from Eq. (\ref{18}) by a partial trace over
the photons, i.e. by
\begin{equation}\label{20}
| 0_{ph} \rangle \left( \hbox{tr}_{ph} P_1 \rho(t_{n + 1}) P_1 \right)
\langle 0_{ph} | / {\rm tr} (\cdot)~.
\end{equation}
The physical reason for this is that for the atomic description alone
it should make no difference in infinite space whether or not the
photons are absorbed, as long as they are sufficiently far away from
the atom and no longer interacting with it \cite{cav}.
Eq. (\ref{20}) can be calculated by perturbation theory for
$U_I(t_{n + 1}, t_n)$, as in Section 2. Now, the first-order
contribution suffices and one obtains in a straightforward way for
the reset state
\begin{eqnarray}
\hat{\cal {R}} (\Delta t) \rho_A (t_n) &\equiv& {\rm tr}_{ph} P_1 U(t_{n +
1}, t_n) | 0_{ph} \rangle \rho_A (t_n) \langle 0_{ph} | U (t_{n +
1}, t_n)^* P_1 \nonumber\\
 & = & e^{-i H^0_A \Delta t / \hbar} \sum_{{i j \ell m \atop i > j, \ell >
m}} |j \rangle \langle i
| \rho_A (t_n) | \ell \rangle \langle m | e^{i H^0_A \Delta t /
\hbar} \nonumber \\
&\times& \int^{\Delta t}_0 dt' \int^{\Delta t}_0 dt'' e^{i \omega_{\bf k}(t'
 - t'')
- i \omega_{ij} t' +i \omega_{\ell m} t''}
\sum_{{\bf k} \lambda} \frac{e^2 \omega_{\bf k}}{2 \varepsilon_0
\hbar V} ({\bf D}_{j i} \cdot {\bf \varepsilon}_{{\bf k}
\lambda})({\bf \varepsilon}_{{\bf k} \lambda} \cdot {\bf D}_{\ell m})~.
\nonumber\\
\label{21}
\end{eqnarray}

Note that the atomic trace $\hbox{tr}_A \hat{\cal {R}} (\Delta t) \rho_A
(t_n)$ gives the probability for a photon to be found at time $t_{n +
1}$ under the condition that no photons were found  at time $t_n$.

The external field drops out in the first-order contribution since
its action during the short time $\Delta t$ is of second order only.
To apply the Markov property, we decompose the rectangular integration
domain over $t'$ and $t''$ in Eq. (\ref{21}) into two triangles,
leading to \cite{deriv}
\begin{equation}\label{22}
\int^{\Delta t}_0 dt' e^{-i (\omega_{ij} - \omega_{\ell m}) t'}
\int^{t'}_0 dt'' e^{i(\omega_{\bf k} - \omega_{\ell m})(t' - t'')} +
\int^{\Delta t}_0 dt'' e^{-i (\omega_{ij} - \omega_{\ell m})t''}
\int^{t''}_0 dt' e^{-i (\omega_{\bf k} - \omega_{ij})(t'' - t')}~.
\end{equation}
As in Eqs. (\ref{10}) and (\ref{11}) the inner integrals can be
replaced by $\pi \delta (\omega_{\bf k} - \omega_{\ell m})$ and $\pi
\delta (\omega_{\bf k} - \omega_{ij})$, respectively, plus principal
values.
In the limit $V \rightarrow \infty$ the sum over
$\bf k$ becomes an integral over $d^3 k$ as in Eq. (\ref{11a}). With
$\Gamma_{i j k \ell}$ given by Eq. (\ref{12}) we thus obtain
\begin{equation}\label{23}
\hat{\cal {R}} (\Delta t) \rho_A (t_n) = e^{-i H^0_A \Delta t/\hbar}
\sum_{{i j \ell m \atop i > j, \ell > m}}
\left\{ \Gamma_{j i \ell m} + \Gamma_{\ell m j i} \right\} | j
\rangle \langle i | \rho_A (t_n) | \ell \rangle \langle m | e^{i
H^0_A \Delta t / \hbar} \int^{\Delta t}_0 dt' e^{- i (\omega_{ij} -
\omega_{\ell m}) t'}
\end{equation}

Up to normalization this is the state of an atom after a photon
detection at time $t_n + \Delta t$, under the condition that no
photons were found at time $t_n$.
Its trace gives the probability for this event.

Even after normalization the state will in general depend in an
oscillatory way on $\Delta t$, except in two limiting cases which
were discussed in Ref. \cite{reset}. If two optical transition
frequencies $\omega_{ij}$ and $\omega_{\ell m}$ are far apart, then
$| \omega_{ij} - \omega_{\ell m} | \Delta t \gg 1$ and the integral
in Eq. (\ref{23}) vanishes. The other case is when two optical
transition frequencies are very close so that $| \omega_{i j} -
\omega_{\ell m} | \Delta t \ll 1$. Then the integral in Eq.
(\ref{23}) is essentially 1, and in this case one may obtain
interesting coherence effects which were discussed for the $\Lambda$
system in Ref. \cite{HePl3}.

There is a close connection between the reset operator, the
conditional Hamiltonian, and the Bloch equations. By Eq. (\ref{19})
one can write Eq. (\ref{20}) for the reset state as
\begin{equation}
{\rm tr}_{ph} P_1 \rho(t_{n + 1}) =  {\rm tr}_{ph} U(t_{n+1},t_n)\rho(t_n)
U(t_{n+1},t_n)^*
- \langle 0_{ph} | U(t_{n + 1}, t_n)\rho(t_n) U(t_{n + 1}, t_n)^*
| 0_{ph} \rangle
\label{23a}
\end{equation}
with $\rho (t_n) = | 0_{ph} \rangle\rho_A(t_n) \langle 0_{ph} |$. The
first term on the right hand side is the definition of
the atomic density matrix in the Bloch equations, while the second
is given by the conditional time development operator. Hence one can
obtain the reset operator also from a knowledge of the Bloch
equations and $H_{\rm cond}$.

For the general case the reset operator in Eq. (\ref{23})
is cumbersome to work with, and it will now be shown that one can
replace it by a refined  -- and simpler -- expression which gives
equivalent results for all photon counting questions.

We define the atomic superoperator $\hat{J}$ by
\begin{equation}\label{24}
 \hat{J} \rho_A \equiv \sum_{{i j \ell m \atop i > j, \ell > m}}
 \left\{ \Gamma_{j i \ell m} + \Gamma_{\ell m j i} \right\} | j
\rangle \langle i | \rho_A | \ell \rangle \langle m |~.
\end{equation}
{\em Formally} one has
\begin{equation}\label{25}
\hat{J} = \lim_{\Delta t \rightarrow 0} \hat{\cal {R}} (\Delta
t) / \Delta t
\end{equation}
but of course this limit is not really physically allowed. However, the
photon counting probabilities obtained from $\hat{J}$ for physically
allowed $\Delta t'$s turn out to be in agreement with those obtained
from $\hat{\cal {R}} (\Delta t)$.

As in Ref. \cite{reset} we define the atomic superoperator
$\hat{S} (t, t_0)$ by
\begin{equation}\label{29}
\hat{S} (t, t_0) \rho_A \equiv
U_{\rm cond}(t,t_0 ) \rho_A U_{\rm cond}(t,t_0)^* \,.
\end{equation}
Based on $\hat{J}$ as reset operator the probability density $w
(\tau_1, \cdots , \tau_k; [0, t])$ for finding a photon exactly at
the times $\tau_1, \cdots, \tau_k$ in $[0, t]$ is given by \cite{reset}
\begin{equation}\label{26}
w (\tau_1, \cdots, \tau_k; [0, t]) = {\rm tr}_A (\hat{S} (t, \tau_k)
\hat{J} \hat{S} (\tau_k, \tau_{k - 1}) \hat{J} \cdots \hat{J} \hat{S}
(\tau_1, 0_1) \rho_A(0) )
\end{equation}
To see how this compares with the photon counting probabilities
obtained from $\hat{\cal {R}} (\Delta t)$ we first express the latter
by means of $\hat{J}$ as follows. Let $\hat{T}_0 (t, t_0)$ be the
superoperator for the {\em free} time evolution of atomic density
matrices,
\begin{equation}\label{27}
\hat{T}_0 (t, t_0) \rho_A \equiv e^{-i H^0_A (t- t_0)/\hbar} \rho_A
e^{iH^0_A (t - t_0)/\hbar}
\end{equation}
and let $\hat{T}(t, t_0)$ be the time evolution including driving and
damping, i.e. given by the solution of the Bloch equations. With
$\hat{T}_0$ Eq. (\ref{23}) can obviously be written as
\begin{equation}\label{28}
\hat{\cal {R}} (\Delta t) = \int^{\Delta t}_0 dt' \hat{T}_0 (\Delta t -
t',0) \hat{J} \hat{T}_0 (t', 0) ~.
\end{equation}

We now use this expression for $\hat{\cal {R}} (\Delta t)$ to
calculate the probability of finding a photon in the small time
interval $[t_i, t_{i + 1}]$ and
none between $0$ and $t_i$ and none between $t_{i + 1}$ and $t$. With a
simple change of integration range in Eq. (\ref{28})
the desired probability is given by
\begin{equation}\label{30}
{\rm tr}_A \hat{S} (t, t_{i + 1}) \hat{\cal {R}} (\Delta t) \hat{S}
(t_i, 0) \rho_A(0) = \int^{t_{i+1}}_{t_i} d \tau_1 {\rm tr}_A \hat{S}
(t, t_{i+1}) \hat{T}_0 (t_{i+1}, \tau_1) \hat{J} \hat{T}_0 (\tau_1,
t_i) \hat{S} (t_i, 0) \rho_A(0)~.
\end{equation}
Now, since $\Delta t = t_{n + 1} - t_n$ is small compared to the
inverse damping and driving, one may -- during $\Delta t$ -- replace
$\hat T_0$ by $\hat{T}$ or by $\hat{S}$. Here we replace it by $\hat{S}$
and obtain for Eq. (\ref{30})
\begin{equation}\label{31}
\int^{t_{i + 1}}_{t_i} d \tau_1 {\rm tr}_A (\hat{S} (t, \tau_1) \hat{J}
\hat{S} (\tau_1, 0) \rho_A(0))~.
\end{equation}
The integrand is just the probability density $w(\tau_1; [0,t])$ of
Eq. (\ref{26}) associated with $\hat{J}$ for $k = 1$. It is clear that this
argument immediately carries over to more than a single photon
detection. The essential point is seen to be the fact that during a
single time interval $\Delta t$ the damping has negligible effect.

We thus have shown that for all photon countings we can use the
refined -- or idealized -- reset operator $\hat{J}$ and the densities
of Eq. (\ref{26}).

\vspace*{1cm}

\noindent {\bf 4. Connection with Bloch Equations}

\vspace*{0.5cm}

\noindent In Ref. \cite{reset} it had been shown that an ensemble of
atoms whose individual time development proceeded according to
$H_{{\rm cond}}$ and the reset operator $\hat{J}$ after photon
detections (``jumps") also obeys the Bloch equations. If instead one
uses the reset operator $\hat{\cal {R}} (\Delta t)$ from Eq.
(\ref{23}), to what extent do the Bloch equations still hold?

The subensemble of atoms with no photons until $t$ is described by
$\hat{S} (t, 0) \rho_A(0)$ and the subensemble with last photon
detection before $t$ at time $t_i$ by
\[
\hat{S}(t, t_i) \hat{\cal {R}} (\Delta t) \rho_A (t_{i-1})
\]
where $t = n \Delta t$ and $t_i = i \Delta t$. Thus the complete
ensemble of atoms is described at time $t$ by
\[
\rho_A (t) = \hat{S} (t, 0) \rho_A(0) + \sum^{n - 1}_{i = 1} \hat{S}
(t, t_i) \hat{\cal {R}} (\Delta t) \rho_A (t_{i - 1})~.
\]
With Eq. (\ref{28}) and a change of integration variable one obtains
\begin{equation}\label{32}
\rho_A (t) = \hat{S} (t, 0) \rho_A (0) + \sum^{n-1}_{i=1}
\int^{t_n+1}_{t_i} d \tau \hat{S}(t, t_i) \hat{T}_0 (t_i, \tau)
\hat{J} \hat{T}_0 (\tau, t_{i-1}) \rho_A (t_{i-1})~.
\end{equation}
As noticed before, during a single time interval $\Delta t$ the
driving and damping has negligible effect, and therefore one can
replace in Eq. (\ref{32}), for this short time interval, $\hat{T}_0
(t_i, \tau)$ by $\hat{S}(t_i, \tau)$ and $\hat{T}_0 (\tau, t_{i-1})$ by
$\hat{T} (\tau, t_{i-1})$, where the latter is
the time development operator of the Bloch equations including
driving and damping. Since $\hat{T} (\tau,
t_{i-1}) \rho_A (t_{i-1}) = \rho_A (\tau)$, one obtains with this
approximation
\[
\rho_A(t) = \hat{S}(t, 0) \rho_A(0) + \sum^{n-1}_{i=1}
\int^{t_{i+1}}_{t_i} d \tau \hat{S} (t, \tau) \hat{J} \rho_A(\tau)
\]
and thus
\begin{equation}\label{33}
\rho_A(t) = \hat{S}(t, 0) \rho_A (0) + \int^t_0 d \tau \hat{S}(t,
\tau) \hat{J} \rho_A (\tau)
\end{equation}
which is just the equation one would have obtained with $\hat{J}$ as
reset operator. Introducing a coarse-grained time scale one can now
differentiate with respect to $t$ and obtains
\[
\dot{\rho}_A = -\frac{i}{\hbar} \left\{ H_{{\rm cond}} \rho_A (t) -
\rho_A (t) H^*_{{\rm cond}} \right\} + \hat{J} \rho_A (t)
\]
which agrees with the usual Bloch equations \cite{PP}.\\[1cm]

\noindent {\bf 5. Discussion}

\vspace*{0.5cm}
The idea of the quantum jump approach is to describe the
radiating atom between photon detections by a conditional (or
reduced)
time evolution operator giving the time development under the
condition that no photon has been detected. After a photon
detection one has to reset the atom to the reset state ("jump"),
with ensuing
reduced time development, and so on.
For a driven  system with many emissions one then obtains a
stochastic path ("quantum trajectory"). An ensemble of such paths for
an ensemble of driven atoms satisfies the  Bloch equations.
In fact both  quantum trajectories and Bloch
equations are possible and equivalent ways to describe the time evolution
of an ensemble of  fluorescing atoms, but the former is also easy to
apply to the emission behavior of a single atom.

As outlined above the photon measurements are taken at rapidly
repeated discrete times. On a coarse-grained time-scale the paths
can be regarded as continuous, and the quantum jump approach can be
considered as a simple model for continuous measurements. The photon
operators $\hat S$ and $\hat J$ for the photon counting distributions
of Sections 3 in general
not only satisfy all the requirements of the axiomatic
continuous-measurement theory of Ref. \cite{DS}, but are explicitly
given, in contrast to the axiomatic theory where they have to be more
or less guessed. In some cases positivity problems for $\hat S$ and $\hat J$
can arise, as discussed further below.

In Ref. \cite{PP} $ \|P_0 U(t,0)|0_{ph}\rangle|\psi\rangle\|^2 $,
the no-photon probability {\em at}  time $t$  was
calculated, i.e.  $= \|\langle 0_{ph} | U (t, 0) | 0_{ph} \rangle | \psi
\rangle\|^2$,
which is, in principle, different from the probability of finding no
photon {\em until} time $t$, since in the latter case
one has to measure in between.
Interestingly, though, Ref. \cite{PP} finds for the no-photon
probability {\em at} time $t$ the same expressions we
do for the probability {\em until} time $t$,
except for their use of ${\bf p}\cdot{\bf A}$ coupling \cite{cav}.
This seems
to indicate that the reductions used in Eq. (\ref{4}) are closely
related to the Markov assumption and can be replaced by it. Ref.
\cite{Reibold} derives photon counting statistics by an approach which
can be regarded as an alternative to that of Ref.
\cite{HeWi,reset} and which uses the projector formalism together
with the Markov property, considering only finite macroscopic time
intervals. Interruptions by numerous hypothetical measurements are
not needed, and neither is the notion of continuous measurements.

The approximations used in the previous sections are the same as those
employed in the usual derivations of the Bloch equations \cite{Mi,PP}, in
particular  the use of the Markov property, i.e. the fact
that the correlation function $\kappa(\tau)$  is very
narrowly peaked and drops off rapidly. This is a standard assumption
in the derivation of Bloch equations \cite{Mi,PP,CT}. The integration
of the approximated time derivatives over $t'$ from $t_i$ to $t_{i +
1}$ in Eq. (\ref{12a}) introduces an error for small
$t' - t_i$, just as in the derivation Bloch equations \cite{Reib},
and a similar error occurs in the reset operator.
The omission of the principal values in Eqs. (\ref{17}) and
(\ref{24}) is not necessary. If one retains them, as for example in
Ref. \cite{Mi}, only the form of the $\Gamma_{ij \ell m}$'s
will be changed. The conditional Hamiltonian will then acquire an
additional term through $\Gamma$, and Eq. (\ref{17}) should be
written in terms of the hermitian (real) part $\Gamma_{\rm r}$ and
antihermitian (imaginary) part $\Gamma_{\rm i}$ as
\begin{equation}\label{D1}
H_{{\rm cond}} = H_A(t) + \Gamma_{\rm i} - i \Gamma_{\rm r}~.
\end{equation}
The imaginary part of $\Gamma$ then contributes to the level shifts,
but the effect of its possible non-diagonal parts deserves closer
examination, both for the quantum jump approach and for the Bloch
equations.

 As seen from Eqs. (\ref{20}) and (\ref{21}), the reset
superoperator must preserve positivity, i.e. must map positive
operators onto positive operators. The question is whether, after the
approximations used in its derivation, this property still holds.
Without explicit calculation of the principal value terms in
$\Gamma_{i j \ell m}$ little can be said. However, if one drops the
principal value terms, as in the Bloch equations of Ref. \cite{PP},
then positivity  may be lost. In fact, for parallel transition dipole
moments  the reset matrix of the $\Lambda$ system \cite{Ag} calculated from
Eq. (\ref{24}) then has a tiny negative part for small level
separation.

Similarly, if the transition to a continuous coarse-grained time
scale for the conditional no-photon time development is possible, then
Eq. (\ref{17b}) for the probability density of the first photon shows
that the hermitian (real) part $\Gamma_r$ of $\Gamma$
should be  a positive operator. Without explicit calculation of the principle
value part little can be said, but if these are omitted then
positivity of $\Gamma_{\rm r}$
is lost in some cases when $\Gamma$ has non-diagonal
elements, e.g. for a V system with parallel transition dipole
moments.

These deviations from positivity are probably small in practical
applications. But it would be preferable to work with expressions
which do respect positivity. One way to arrive at such expressions
is to note that in Eq. (\ref{24}) for $\hat{J}$ one has, with
omission of the principal value terms in $\Gamma_{i j \ell m}$,
\begin{equation}\label{33b}
\Gamma_{j i \ell m} + \Gamma_{\ell m j i} \sim {\bf D}_{j i} \cdot
{\bf D}_{\ell m} \frac{1}{2}
\left( \omega_{i j}^3 + \omega_{\ell m}^3 \right)~.
\end{equation}
Now, if one replaces the arithmetic mean $\frac{1}{2} \left(
\omega_{i j}^3 + \omega_{\ell m}^3 \right)$ by the geometric mean
$\sqrt{\omega_{i j}^3 \omega_{\ell m}^3}$ then it is easily seen that
positivity of the resulting reset operator is automatic.

Since $\hbox{tr} \hat{J} \rho_A$  gives the probability density for the
first photon, one has, by Eq. (\ref{17b}),
\begin{equation}\label{33c}
\hbox{tr} (\hat{J} \rho_A) = \hbox{tr} (\Gamma + \Gamma^* ) \rho_A
\end{equation}
if the coarse-grained time scale can be used. Once one has obtained a
positivity preserving $\hat{J}$ one can use Eq. (\ref{33c}) to find a
positive $\Gamma_{{\rm r}}$. A systematic procedure to obtain a positivity
preserving expression for the reset operator would be desirable.

In view of the close connection of the quantum jump approach with the
Bloch equations it stands to reason that the possible non-positivity
of the reset matrix and the damping operator could have its
counterpart in the Bloch equations. Indeed, at least if one omits the
principal value parts, as in Ref. \cite{PP}, this can happen for very
short times.
E.g.,  for the $\Lambda$ system with no external driving field
and with parallel transition dipole
moments one can start initially in  the upper state which then,
under the time evolution under the Bloch equations, develops a
small negative part for  times of the order of the inverse optical
frequencies or less, but regains positivity for times larger than that.
We have also found that a similar phenomenon happens
for a V system with parallel transition dipole moments. Positivity in
the Bloch equations is ensured if one make s the replacement of the
arithmetic mean in Eq. (\ref{33b}) by the geometric mean. Although
this may mean in some cases a substantial change of some coefficients
in the Bloch equations, the effect on the behavior of the solutions
is expected to be minimal. A similar non-positivity has been found in
Ref. \cite{Haake}.

This seems to be similar to the effect of the
rotating-wave approximation. Keeping or not keeping the rapidly
rotating terms makes some coefficients considerably different, but
has little effect on the solutions. For the same reason rapidly
oscillatory terms arising through contributions from $\Gamma_{ijj\ell}$
for large $\omega_{j\ell}$ can be omitted. We have kept them here
only to exhibit the symmetry of the formulas.

The derivations in this paper were based on the ${\bf E} \cdot {\bf D}$
coupling. If one uses the ${\bf p} \cdot {\bf A}$ coupling the
results are the same except that the expression for the $\Gamma_{i j
\ell m}$ are slightly different. If one again replaces the arithmetic
by the geometric mean the results become the same.

In conclusion one may say that the degree of reliability of our
results for the conditional (reduced) Hamiltonian and the
reset operator is the same as that of the corresponding Bloch equations.
Because the times involves are very short -- in fact of the order of the
correlation time used in the Markov approximation -- and the non-positivity
very small, there are probably no practical consequences.


\begin{thebibliography}{99}
\bibitem{width} For a simplified case one can see this directly as
follows. The $\omega^2$ inherent in $d^3k$ and the $\omega_k$ in
Eq.(\ref{10a}) give a factor of $\omega^3$. If this $\omega^3$ is
omitted in the definition of $\kappa_{ij\ell m}$ then the result can
be seen by a straightforward calculation of the double integral in
Eq. (\ref{10}). The general case can be reduced to this by
partial integration. We note that for $\omega_{\ell m}$ in the
microwave range the condition $t'- t_i \gg \omega_{\ell m}$ does not
hold. However, the radiative coupling of such levels is extremely
small and is usually neglected in applications.
\bibitem{cav} For a cavity,
however, where photons can return and revivals can occur, the results
are different.
\bibitem{deriv} This decomposition results automatically if one
considers the time derivative of Eq. (\ref{21}), similar as in
Section 2.
\end{thebibliography}
\end{document}